\documentstyle[prd,preprint,aps]{revtex}
\begin{document}
\draft
\title{Heavy boson production through the collision of an ultrahigh-energy
neutrino on a target nucleon}
\author{R.M. Garc\'{\i}a-Hidalgo and A. Rosado}
\address{Instituto de F\'{\i}sica , Universidad Aut\'onoma de Puebla.\\
Apartado Postal J-48, Colonia San Manuel, Puebla, Puebla. 72570, M\'exico}
\date{\today}
\maketitle

\begin{abstract}
We discuss $W$ and $Z$ production through the deep inelastic $\nu_l{\cal N}$-
scattering in the context of the standard model $SU(3)_C \times SU(2)_L
\times U(1)$ of the strong and electroweak interactions. We find the cross
section rates for the process $\nu_l + {\cal N} \rightarrow l^- + W^+ +
X$ for the case of ultrahigh-energy neutrinos ($10^{14} \, eV \leq E_{\nu}
\leq 10^{17} \, eV$) colliding on a target nucleon ($E_{\cal N}=m_{\cal N}$).
We also calculate $\sigma ( \nu_l + {\cal N} \rightarrow l^- + X)$ in order
to compare it with $\sigma ( \nu_l + {\cal N} \rightarrow l^- + W^+ + X)$.
\end{abstract}
 
\narrowtext

\section{Introduction}

\par Large-scale neutrino telescopes \cite{telescopes} have as a main goal
the detection of ultrahigh-energy (UHE) cosmic neutrinos ($E_\nu \geq
10^{12}$ $eV$) produced outside the atmosphere (neutrinos produced by
galactic cosmic rays interacting with interstellar gas, and extragalactic
neutrinos) \cite{uhe-neut}. UHE neutrinos can be detected by observing
long-range muons produced in charged-current neutrino-nucleon interactions.
A very enlightening discussion on UHE neutrino interactions is given by R.
Gandhi {\it et al.} \cite{gandhi}.

\par The detection of UHE neutrinos will provide us with the possibility to
observe $\nu {\cal N}$-collisions with a neutrino energy in the range
$10^{12} \, eV \leq E_\nu \leq 10^{21} \, eV$ and a target nucleon at rest.
Earlier estimations of the $W$ production rates in $\nu {\cal N}$-scattering
were given in 1970, by H. H. Chen \cite{chen}, and by J. Reiff \cite{reiff},
in 1971 by R. W. Brown and J. Smith \cite{brown1}, and by F. A. Berends and
G. B. West \cite{berends}. In those works the calculations were made using
neither the standard model \cite{stanmod} nor the parton model
\cite{partmod}. One of the aims of detecting UHE neutrinos is to
look for physics beyond \cite{instantons} the standard model \cite{stanmod}.
However, the event rates for most of the exotic processes could be of the
same order as the event rates for heavy gauge boson production in
$\nu {\cal N}$-scattering. Hence, even though the cross sections for heavy
boson production are expected not to be large enough to allow for detailed
investigations of the heavy bosons properties, it is necessary to calculate
them. Only if heavy boson production, which could be an important background
for physics beyond the standard model, is completely known and subtracted
from the exotic event rates, the new physics could be investigated. Therefore,
we discuss in this paper heavy vector boson production in the deep inelastic
$\nu_l {\cal N}$-scattering within the frame of the standard model $SU(3)_C
\times SU(2)_L \times U(1)$ of the strong and electroweak interactions and
using the parton model \cite{partmod} with the parton distribution functions
reported by J. Pumplin {\it et al.} \cite{pumplin}. We use the CTEQ PDFs
provided in a $n_f=5$ active flavours scheme.

\par The cross section for the process $\nu_l + {\cal N} \rightarrow l^- +
W^+ + X$ is expected to be the most important one because it gets
contributions from photon exchange diagrams (see figs. 2.1-2.5). We thus
present results for the total cross section for $W^+$ production through the
processes $\nu_l + {\cal N} \rightarrow l^- + W^+ + X$ as a function of the
neutrino energy in the range $10^{14} \, eV \leq E_{\nu} \leq 10^{17} \, eV$
\footnote{We take $E_{\nu} \leq 10^{17} \, eV$, in order to use the CTEQ PDFs
in the range of the ($x', Q'{}^2$) parameter space in which these have been
found to be valid: $10^{-6} \leq x' \leq 1$ and $1.3 \, GeV \leq
\sqrt{Q'{}^2} \leq 10^4 \, GeV$} and taking $E_{\cal N}=m_{\cal N}$ for the
nucleon. We also calculate $\sigma ( \nu_l + {\cal N} \rightarrow l^- + X)$
to find out how it compares to $\sigma ( \nu_l + {\cal N} \rightarrow l^- +
W^+ + X)$.

\par This paper is organized as follows. In section II, we provide the 
expressions for the kinematics of the production of a massive vector bosons
in deep inelastic neutrino-nucleon scattering and discuss the influence of
setting cuts on the phase space of the produced particles. In section III, we
perform the explicit calculation of the matrix elements and the differential
cross section for polarized scattering. We present and discuss our results
for the total cross section for the case of $W^+$ production in section
IV. Finally, section V contains a summary of our conclusions.

\section{Kinematics}

\par In this section we will discuss the kinematics and phase space of the
production of a heavy vector boson $B = (W^\pm, Z^0)$ through the inclusive
processes
\begin{equation}
\nu_l + {\cal N} \rightarrow  l' + B + X
\end{equation}
\noindent where $\nu_l, \, {\cal N}, \, l'$ and $B$ stand for the incoming
neutrino, the target nucleon, the final lepton ($\nu_l$ or $l^-$) and the
produced massive boson, respectively. We will denote the four-momenta of
these particles by $p$, $P_{\cal N}$, $p'$ and $k$, respectively. In
accordance with the kinematics for the collision of a neutrino on a target
nucleon, the following construction is chosen:
\begin{equation}
p^\mu=E_{\nu}(1,0,0,1), \hspace{2.5cm} P^\mu_{\cal N}=m_{\cal N}(1,0,0,0).\\
\end{equation}

\par As usual we define the invariants:
\begin{equation}
\begin{array}{lll}
s&=&(p+P_{\cal N})^2 \\
Q^2&=&-(p-p')^2 \\
\nu&=&P_{\cal N}(p-p') \\
s'&=&(p+P_{\cal N}-k)^2 \\
Q'{}^2&=&-(p-p'-k)^2 \\
\nu'&=&P_{\cal N}(p-p'-k) \\
\end{array}
\end{equation}
\noindent and the dimensionless variables:
\begin{equation}
x=\displaystyle\frac{Q^2}{2\nu}, \hspace{1cm}
y=\displaystyle\frac{2\nu}{s}, \hspace{1cm}
\tau=\displaystyle\frac{s'}{s}, \hspace{1cm}
x'=\displaystyle\frac{Q'{}^2}{2\nu'}, \hspace{1cm}
y'=\displaystyle\frac{2\nu'}{s}.
\end{equation}
\noindent The physical region of these kinematical variables is obtained by
requiring that the scalar products of any two particle four-momenta be
positive, the determinants $\Delta_3$ of any three particle four-momenta
positive, and the determinant $\Delta_4$ of the four independent four-momenta
(whenever possible we will neglect the nucleon and lepton masses):
\begin{equation}
\nonumber\\[1mm]
\Delta_4(p,p',P_{\cal N},k)=
\left|
\begin{array}{cccc}
0 & pp' & pP_{\cal N} & pk \\
p'p & 0 & p'P_{\cal N} & p'k \\
P_{\cal N}p & P_{\cal N}p' & 0 & P_{\cal N}k \\
kp & kp' & kP_{\cal N} & M^2_B \\
\end{array}
\right|
\nonumber\\[1mm]
\end{equation}
\noindent negative \cite{Kinematics}.

\par From the positivity of the scalar products we  find:
\begin{equation}
0 \leq x < x' \leq 1,\hspace{0.75in} 0 \leq y'< y \leq 1.
\end{equation}
Explicit evaluation of $\Delta_4$ using (3), (4) and (5) gives:
\begin{equation}\
\Delta_4 = - (s/2)^4 x'{}^2 y^2 (\tau_b - \tau)(\tau - \tau_a)
\end{equation}
where
\begin{equation}
\begin{array}{lll}
\tau_{a,b}&=&(1-x')(1+y'-y)+(1/y)\{(1-y)[(x'-x)(y-y')- \mu]+xy'\nonumber\\
        & &                           \mp \, 2\sqrt{xy'(1-y)[(x'-x)(y-y')-\mu]} \, \}.
\end{array}
\end{equation}
\noindent with $\mu=M^2_B/s$ and $s=2 m_{\cal N} E_{\nu}$.
The condition $\Delta_4 \leq 0$ can be used to restrict $\tau$:
\begin{equation}
\tau_a \leq \tau \leq \tau_b,
\end{equation}
From the requirement that $\tau_{a,b}$ be real, together with $x\geq0$,
$y'\geq0$ and $y\leq1$, there follows the inequality
\begin{equation}
(x'-x)(y-y') \geq \mu,
\end{equation}
\noindent which is equivalent to $\Delta_3(p,p',k) \geq 0$.
Inequality (10) can be used to replace (6) by the more stringent relations:
\begin{equation}
\begin{array}{lll}
0 \leq x \leq 1-\mu/y, \hspace{0.75in} x+\mu/y \leq x' \leq 1,\nonumber\\
\mu \leq y \leq 1, \hspace{0.75in} 0 \leq y' < y-\mu/(x-x').
\end{array}
\end{equation}
\noindent The expressions given in (8), (9) and (11) define the physical
region for the dimensionless variables $x$, $x'$, $y$, $y'$ and $\tau$.
We have taken all fermion masses to be zero, this implies that in the 
calculation of the total cross section the integration over the momentum 
transfer square extends up to zero. Hence, we have no natural regulator for 
the contribution of the photon-exchange diagrams. This is one of the reasons 
why we need cuts on $Q^2$ and $Q'{}^2$. Another reason is that the parton 
distributions can be used only when $Q^2$ and $Q'{}^2$ are not too small. 
Furthermore, in order to separate deep inelastic from elastic scattering, a
cut on the invariant mass $W$ of the unobserved particles in the final state
is required. Therefore, besides the kinematical conditions (8), (9), (11) 
we have in general also the following:
\begin{eqnarray}
Q^2&=&sxy\geq Q^2_{c},\nonumber\\
Q'{}^2&=&sx'y'\geq Q'{}^2_{c},\\
W&=&sy'(1-x')+m_{\cal N}^2\geq W_{c}.\nonumber
\end{eqnarray}
\noindent The cuts for $Q^2$, $Q'{}^2$ and $W$ constrain further the
physically allowed region for the process (1). This region can now be
written in terms of the dimensionless variables as follows:
\begin{eqnarray}
y_a \leq y \leq y_b, \hspace{0.75in}& x_a \leq x \leq x_b,\nonumber \\
x'_a \leq x' \leq x'_b, \hspace{0.75in}& y'_a \leq y' \leq y'_b.
\end{eqnarray}
with
\begin{eqnarray}
\vspace{3mm} 
y_a&=&\max\{(M_B+\sqrt{W_{c}} \, )^2/s,\,(M^2_B+Q^2_{c})/s\},
\hspace{0.75in} y_b=1, \vspace{3mm} \nonumber\\
x_a&=&Q^2_{c}/sy,\hspace{0.75in} x_b=1-(M_B+\sqrt{W_{c}} \, )^2/sy,
\vspace{3mm} \nonumber\\
x'_a&=&\max\{\displaystyle\frac{1}{2y}[y(1+x)+(M^2_B-W_{c})/s\:
-\sqrt{[y(1-x)-(M^2_B+W_{c})/s]^2-4M^2_BW_{c}/s^2} \, ], \vspace{3mm} \nonumber\\
    & &\hspace{.4in} \displaystyle\frac{1}{2y}[xy+(M^2_B+Q'{}^2_{c})/s\:
+\sqrt{[xy+(M^2_B+Q'{}^2_{c})/s]^2-4xyQ'{}^2_{c}/s} \, ] \}, \vspace{3mm} \nonumber\\
x'_b&=&\displaystyle\frac{1}{2y}\{y(1+x)+(M^2_B-W_{c})/s\:
+\sqrt{[y(1-x)-(M^2_B+W_{c})/s]^2-4M^2_BW_{c}/s^2} \, \}, \vspace{3mm} \nonumber\\
y'_a&=&\max\{W_{c}/s(1-x'),\,Q'{}^2_{c}/sx'\},\hspace{0.75in} 
y'_b=y-M^2_B/s(x'-x).\nonumber 
\end{eqnarray} 
\noindent The limits for $\tau$ are not modified by the cuts given in (12),
thus remaining as in (9).

\section{The differential cross section for inclusive boson production}

The differential cross section $d\sigma^{\nu_l{\cal N}}$ for (1) is
calculated in the parton model from the cross section $d\sigma^{\nu_l q}$ of
the parton subprocess
\begin{equation}
\nu_l+q\rightarrow l'+q'+B
\end{equation}

\noindent (where $\nu_l$, $q$, $l'$, $q'$ and $B$ stand for the incoming
neutrino, a quark inside the target nucleon, the final lepton ($\nu_l$ or
$l^-$), the outgoing quark and the produced massive boson) and the parton
distribution functions $f_q(x^*,\tilde{Q^2})$, which are the probabilities to
find a quark $q$ with the fraction $x^*$ of the nucleon momentum: $q^\mu=
x^* P^\mu_{\cal N}$, in a scattering process with momentum transfer square
$\tilde{Q^2}$. We will denote the four-momenta of these particles by $p$,
$q$, $p'$, $q'$ and $k$, respectively. As usual, we define for the parton
process the invariant variables $\hat{s},\,\hat{Q}^2,\,\ldots,\,\hat{\nu}'$
and using the definitions given in (3) we get the following relations:
\begin{equation} 
\begin{array}{lllll}
\hat{s}&=&(p+q)^2&=&x^* s,\\
\hat{Q}^2&=&-(p-p')^2&=&Q^2,\\
\hat{\nu}&=&q(p-p')&=&x^* \nu,\\
\hat{s}'&=&(p+q-k)^2&=&s'-(1-x^*)s+2(1-x^*)(\nu-\nu'),\\
\hat{Q}'{}^2&=&-(p-p'-k)^2&=&Q'{}^2,\\
\hat{\nu}'&=&q(p-p'-k)&=&x^* \nu',
\end{array}
\end{equation} 
The variables $\hat{s},\,\ldots,\,\hat{\nu}'$ are not independent. For
massless partons we have $\hat{Q}'{}^2=2\hat{\nu}'$ and consequently 
$x^*=Q'{}^2/2\nu'$. Comparing with the definition of $x'$ given in (4) we
conclude that $x'=x^*$, {\it i. e.} $q^\mu=x' P^\mu_{\cal N}$.

The parton cross section is obtained from the invariant matrix element
${\cal M}(\nu_l q\rightarrow l'q'B)$:
\begin{equation}
d\sigma^{\nu_l q}=\frac{(2\pi)^{-
5}}{2 \, \hat{s}}\,\frac{1}{4}\,\sum_{s,\lambda} \left|{\cal M}^{\nu_l q}
\right|^2 \,d\Gamma_3\,.
\end{equation} 
The sum is performed over all fermion spin states $s$ and the polarizations
$\lambda$ of the produced boson. With help of the different sets of variables
introduced in (15), (3) and (4), and by using (7), the 3-particle phase space
$d\Gamma_3$ can be expressed as follows:
\begin{eqnarray}
d\Gamma_3&=&\frac{d^3p'}{2E_{p'}}\,\frac{d^3q'}{2E_{q'}}\,\frac{d^3k'}{2E_k}
\delta^4 (p + q - p' - q' - k) \nonumber\\
&=&\frac{\pi}{8\hat{s}}\,\frac{d\hat{Q}^2d\,\hat{\nu}d\,\hat{s}'d\,\hat{Q}'{}^2d
\,\hat{\nu}'}{\sqrt{-\Delta_4(p,q,p',k)}}\,\delta(\hat{Q}'{}^2-2\hat{\nu}')\nonumber\\
&=&\frac{\pi}{8s}\,\frac{dQ^2d\nu\,ds'\,dQ'{}^2\,d\nu '}{\sqrt{-
\Delta_4(p,P_{\cal N},p',k)}}\,\delta(Q'{}^2-2x'\nu')\nonumber\\
&=&\frac{\pi s}{8} \frac{dx \, dy \, dy' \, d\tau}{\sqrt{(\tau_b - \tau)
(\tau - \tau_a)}}
\end{eqnarray}
From (16) and (17) we obtain
$$d\sigma^{\nu_l q}=\frac{1}{2^7 \, (2 \pi)^4} 
\,\sum_{s,\lambda} \left|{\cal M}^{\nu_l q}\right|^2\,\frac{1}{x'}\,
\frac{dx \, dy \, dy' \, d\tau}{\sqrt{(\tau_b - \tau)(\tau - \tau_a)}}.$$
The Feynman diagrams for
${\cal M}^{\nu_l q}$ are depicted in fig. 1. The heavy boson can be produced
from the lepton line (fig. 1: a, b), the (anti)quark line (fig. 1: c, d) or
via the non-Abelian vertex from the line of the exchanged bosons (fig. 1: e).
We write
$${\cal M}^{\nu_l q}={\cal M}^{\nu_l q,l}+{\cal M}^{\nu_l q,h}+{\cal M}^{\nu_l q,n}.$$
Diagrams containing the exchange of Higgs bosons have been neglected because
of the smallness of the Higgs fermion coupling. We present the explicit
expressions for these matrix elements in a way which is suited not only for
the standard model of the electroweak interaction but also for extended
models containing more vector bosons than the SM. Therefore, denoting with
$f$ the fermions and with $B$ the gauge bosons, we define the couplings:
$$\bar{\psi}_{f'} \, \psi_f \, B_\mu \, \Rightarrow \, i\gamma_\mu(v^B_{f'f}-
a^B_{f'f}\gamma_5)=i\gamma_\mu(L^B_{f'f}P_L+R^B_{f'f}P_R), $$
and
$$B_\mu(p) \, B'_\nu(p') \, B''_\rho(p'') \, \Rightarrow \, ig_{BB'B''}\{g_{\mu\nu}(p-p')_\rho+g_{\nu\rho}(p'-
p'')_\mu+g_{\mu\rho}(p''-p)_\nu\},$$
\noindent with $\; P_{L,R}=(1 \mp \gamma_5)/2$,   $\; L^B_{f'f}=v^B_{f'f}+a^B_{f'f} \;$,
and   $\; R^B_{f'f}=v^B_{f'f}-a^B_{f'f}$. 

\noindent In the case of the standard model the couplings of the fermions to
the $Z$ and $W$ bosons are given as follows ($e=\sqrt{4\pi \alpha}$):
\begin{eqnarray}
v^{Z}_{f'f}&=&-e\:(T^3_f-2 Q_f sin^2\theta_W)/sin2 \theta_W, \nonumber\\
a^{Z}_{f'f}&=&-e\:T^3_f /sin2 \theta_W, \nonumber\\
v^{W}_{f'f}&=&-e/(2 \sqrt{2}\:sin \theta_W), \nonumber\\
a^{W}_{f'f}&=&-e/(2 \sqrt{2}\:sin \theta_W). \nonumber
\end{eqnarray}
$Q_f$ and $T^3_f$ stand for the
charge and the third component of the isospin of the fermion f. For the three
bosons non-Abelian couplings we have:
\begin{eqnarray}
g_{\gamma W^+ W^-}&=&e, \nonumber\\
g_{Z^0 W^+ W^-}&=&e \: cos \theta_W / sin \theta_W. \nonumber
\end{eqnarray} 
The invariant matrix elements ${\cal M}^{\nu_l q,a}$, $a=\{l,h,n\}$ for the 
production of the boson $B$ (polarization vector 
$\varepsilon^\mu(k,\lambda)$) in the scattering of the neutrino with initial
and the outgoing lepton with final longitudinal polarization $P_{l},P_{l'}$
and the quark with initial and final polarizations $P_{q},P_{q'}$ can be
expressed, taking all fermion masses to be zero (hence $P_{l'}=P_l$ and
$P_{q'}=P_q$) in the following form
$${\cal M}^{\nu_l q,a}_{P_q,P_l}=J_\mu^{h,P_q}f^{r,a}_{P_q,P_l}
{\cal F}_r^{\mu\nu}J_\nu^{l,P_l}$$
where the left-, right-handed currents of the leptons and quarks are 
defined as usual
$$J^{l,P_l}_\mu=\bar{u}(p')\:\gamma_\mu P_l\:u(p),$$
and
$$J^{h,P_q}_\mu=\bar{u}(q')\:\gamma_\mu P_q\:u(p).$$
and the tensors ${\cal F}_r^{\mu\nu}$, $r=\{1,2,3\}$ are defined as 
follows:
\begin{equation}
{\cal F}_1^{\mu\nu}=g^{\mu\nu},\hspace{.5in}
{\cal F}_2^{\mu\nu}=\varepsilon^\mu k^\nu-k^\mu\varepsilon^\nu,\hspace{.5in}
{\cal F}_3^{\mu\nu}=i\varepsilon^{\mu\nu\rho\sigma}\varepsilon_\rho k_\sigma.
\end{equation} 
The invariant functions  $f^{r,a}_{P_q,P_l}$ are built from the 
propagators and couplings in the form: 
$$
\begin{array}{rlrl}
f^{1,l}_{P_q,P_l}=&2\varepsilon p \; C^{l,i}_{P_q,P_l}
+2\varepsilon p' \; C^{l,f}_{P_q,P_l},\hspace{0.5in}
&f^{2,l}_{P_q,P_l}=&C^{l,i}_{P_q,P_l}-C^{l,f}_{P_q,P_l},\vspace{1.5mm}\\
f^{3,l}_{P_q,P_l}=&(-1)^{S_1}(C^{l,i}_{P_q,P_l}+C^{l,f}_{P_q,P_l}),
\vspace{1.5mm}\\
f^{1,h}_{P_q,P_l}=&2\varepsilon q \; C^{h,i}_{P_q,P_l}
+2\varepsilon q' \; C^{h,f}_{P_q,P_l},
&f^{2,h}_{P_q,P_l}=&C^{h,i}_{P_q,P_l}-C^{h,f}_{P_q,P_l},\vspace{1.5mm}\\
f^{3,h}_{P_q,P_l}=&(-1)^{S_2}(C^{h,i}_{P_q,P_l}+C^{h,f}_{P_q,P_l}),\vspace{1.5mm}\\
f^{1,n}_{P_q,P_l}=&2\varepsilon (p-p') \; C^{n,i}_{P_q,P_l},
&f^{2,n}_{P_q,P_l}=&-C^n_{P_q,P_l},\vspace{1.5mm}\\
f^{3,n}_{P_q,P_l}=&0,
\end{array}
$$
with $S_1=0,1,0,1$ and $S_2=0,0,1,1$ for $P_q P_l =LL,\; LR,\;RL,\;RR$, 
respectively, and
\begin{eqnarray}
C^{l,i}_{L,L}&=&\sum_{B',l''} L^{B'}_{l'l''} L^{B}_{l'' \nu_l} 
L^{B'}_{q'q}/[(Q'{}^2+M^2_{B'})(M^2_B-2kp)]\nonumber\\
C^{l,f}_{L,L}&=&\sum_{B',l''} L^{B}_{l'l''} L^{B'}_{l'' \nu_l} 
L^{B'}_{q'q}/[(Q'{}^2+M^2_{B'})(M^2_B+2kp')]\nonumber\\
C^{h,i}_{L,L}&=&\sum_{B',q''} L^{B'}_{\nu_l l'} L^{B'}_{q'q''} 
L^{B}_{q''q}/[(Q^2+M^2_{B'})(M^2_B-2kq)]\\
C^{h,f}_{L,L}&=&\sum_{B',q''} L^{B'}_{\nu_l l'} L^{B}_{q'q} 
L^{B}_{q''q}/[(Q^2+M^2_{B'})(M^2_B+2kq')]\nonumber\\
C^n_{L,L}&=&\sum_{B',B''} L^{B'}_{\nu_l l'} L^{B''}_{q'q} 
g_{B'BB''}/[(Q^2+M^2_{B'})(Q'{}^2+M^2_{B''})].\nonumber
\end{eqnarray}
for left-handed polarization of leptons and quarks. For other polarizations L 
has to be replaced by R in a suitable way in (19).

The next step in the calculation of $d\sigma^{l{\cal N}}_{P_q,P_l}$ is to
square ${\cal M}^{\nu_l q}_{P_q,P_l}$:
\begin{eqnarray}
\left|{\cal M}^{\nu_l q}_{P_q,P_l}\right|^2&=&
\left(\sum_a {\cal M}^{\nu_l q,a}_{P_q,P_l}\right)
\left(\sum_b {\cal M}^{\nu_l q,b}_{P_q,P_l}\right)^*
\nonumber
\end{eqnarray}
and sum over all fermion spin states $s$ and the polarizations $\lambda$ of
the produced boson.

We have
\begin{eqnarray}
\sum_{s,\lambda} {\cal M}^{\nu_l q,a}_{P_q,P_l}
({\cal M}^{\nu_l q,b}_{P_q,P_l})^*&=&
\sum_{s,\lambda,r,r'} J_\mu^{h,P_q}f^{r,a}_{P_q,P_l}{\cal F}_r^{\mu\nu}
J_\nu^{l,P_l}(J{}_{\nu'}^{l,P_l})^*f^{r',b}_{P_q,P_l}
({\cal F}_{r'}^{\mu'\nu'})^*(J_{\mu'}^{h,P_q})^*\nonumber\\
&=&\sum_{\lambda,r,r'} {\cal H}^{P_q}_{\mu\mu'} {\cal L}^{P_l}_{\nu\nu'} 
f^{r,a}_{P_q,P_l} {\cal F}_r^{\mu\nu}f^{r',b}_{P_q,P_l}({\cal F}_{r'}^
{\mu'\nu'})^*\nonumber\\
&=&\sum_{\lambda,r,r'} T^{r,r'}_{P_q,P_l}f^{r,a}_{P_q,P_l}f^{r',b}_{P_q,P_l}
\nonumber
\end{eqnarray} 
with
$${\cal H}_{\mu\nu}^{P_q}={\cal J}_\mu^{h,P_q}{\cal J}_\nu^{h,P_q},\hspace{0.5in}
 {\cal L}_{\mu\nu}^{P_l}={\cal J}_\mu^{l,P_l}{\cal J}_\nu^{l,P_l},$$
and
$$T^{r,r'}_{P_q,P_l}={\cal H}_{\mu\mu'}^{P_q}{\cal F}_r^{\mu\nu}
({\cal F}_{r'}^{\mu'\nu'})^* {\cal L}_{\nu\nu'}^{P_l}.$$
Using the expressions given for ${\cal H}_{\mu\nu}^{P_q}$ and
${\cal L}_{\mu\nu}^{P_q}$ in the appendix and the fact that the
$f^{r,a}_{P_q,P_l},s$ are real functions, it is straightforward to show that
$${\cal M}^{\nu_l q,a}_{P_q,P_l}({\cal M}^{\nu_l q,b}_{P_q,P_l})^*=
({\cal M}^{\nu_l q,a}_{P_q,P_l})^*{\cal M}^{\nu_l q,b}_{P_q,P_l}.$$
Hence
\begin{eqnarray}
\sum_{s,\lambda}\left|{\cal M}^{\nu_l q}_{P_q,P_l}\right|^2&=
      &\sum_{s,\lambda} \left(\left|{\cal M}^{\nu_l q,l}_{P_q,P_l}\right|^2 +
         \left|{\cal M}^{\nu_l q,h}_{P_q,P_l}\right|^2 +
         \left|{\cal M}^{\nu_l q,n}_{P_q,P_l}\right|^2 \right)\nonumber\\
& &+2Re\left\{\sum_{s,\lambda} \left( {\cal M}^{\nu_l q,l}_{P_q,P_l}({\cal M}^{\nu_l q,h}_{P_q,P_l})^* +
     {\cal M}^{\nu_l q,l}_{P_q,P_l}({\cal M}^{\nu_l q,n}_{P_q,P_l})^* +
     {\cal M}^{\nu_l q,h}_{P_q,P_l}({\cal M}^{\nu_l q,n}_{P_q,P_l})^* \right) \right\}\nonumber
\end{eqnarray} 
Explicit expressions for the quantities $T^{r,r'}$ are presented in the
appendix. Also, the explicit summation over the polarizations $\lambda$ of
the produced boson is presented there.

\par A heavy weak boson $B$ can be produced in deep inelastic $\nu_l {\cal N}$
collisions \footnote {the results for $\bar{\nu}_l {\cal N}$-scattering can be
obtained from those of $\nu_l {\cal N}$-scattering by the replacements $\nu_l
\rightarrow \bar{\nu}_l$, $W^{\pm}\rightarrow W^{\mp}$, $l^- \rightarrow l^+$
and $u$-type quarks $\leftrightarrow$ $d$-type quarks} via the following
processes:

\hspace{3cm} $\nu_l + {\cal N}  \rightarrow \nu_l + Z^0 + X,$ \hfill (P.1)   

\hspace{3cm} $\nu_l + {\cal N}  \rightarrow l^- + Z^0 + X,$ \hfill (P.2)

\hspace{3cm} $\nu_l + {\cal N}  \rightarrow \nu_l + W^+ + X,$ \hfill (P.3)

\hspace{3cm} $\nu_l + {\cal N}  \rightarrow \nu_l + W^- + X,$ \hfill (P.4)

\hspace{3cm} $\nu_l + {\cal N}  \rightarrow l^- + W^+ + X,$ \hfill (P.5)

\hspace{3cm} $\nu_l + {\cal N}  \rightarrow l^- + W^- + X,$ \hfill

\noindent The diagrams which contribute in the lowest order at the quark level
to the different reaction mechanisms (production at the leptonic vertex, at
the hadronic vertex and through the boson self interaction) of all these
processes, are depicted in figs. 2.1 - 2.5. The process $\nu_l + {\cal N}
\rightarrow l^- + W^- + X$, is forbidden in the lowest order. We see from
these figures that the reactions (P.1) - (P.4) only get contribution from
heavy boson exchange diagrams and therefore their total cross sections are
expected to be very small and we will not be discussed in this work.

\par The final step in the evaluation of $d\sigma^{\nu_l {\cal N}}$ consists
now in putting together the parton cross sections $d\sigma^{\nu_l q}$ and the
parton distribution functions $f_q(x',\tilde{Q^2})$. In contrast to deep
inelastic $\nu_l {\cal N}$-scattering the choice of $\tilde{Q^2}$ is not
unambiguous in our case since the momentum transfer to the nucleon depends
on the reaction mechanism (in other words, whether the boson is emitted at
the leptonic or at the hadronic vertex). For leptonic production it is
reasonable to take $\tilde{Q}^2=-(p-p'-k)^2=Q'^2$ since $p-p'-k$ is the
momentum transfer to the nucleon. In the case of hadronic production the
obvious choice is $\tilde{Q}^2=-(p-p')^2=Q^2$. For the non-Abelian diagrams a
simple kinematical argument is not sufficient. For heavy boson production in
neutrino quark scattering, unitarity would be violated without a coupling
between the $W$, $Z$ and $\gamma$ bosons. Therefore unitarity is restored
through strong cancellations between these non-Abelian diagrams and either
the leptonic or the hadronic contributions to neutrino lepton scattering.

In this work, we will neglect the contribution from heavy boson exchange
diagrams. Hence, we calculate the cross section of process $\nu_l + {\cal N}
\rightarrow l^- + W^+ + X$ with the following practical prescription which
guarantees the restoration of unitarity
\begin{equation}
d\sigma^{\nu_l{\cal N}}=\sum_q\int dx' f_q(x',Q'{}^2) \left(d\sigma^
{\nu_l q}_{ll} + d\sigma^{\nu_l q}_{ln} + d\sigma^{\nu_l q}_{nn}\right).
\end{equation}
This prescription collects the expressions where the heavy boson is emitted
from the lepton line or the non-Abelian vertex and the interference of these
reaction mechanisms.

\section{Results for $W^+$ production via $\nu {\cal N}$ collisions}

We have already pointed out in the previous section that the process $\nu_l +
{\cal N} \rightarrow l^- + W^+ + X$ is the only one which gets contribution
from photon-exchange diagrams. Therefore, we will restrict ourselves to
calculate numerically the cross section for $W^+$-production through this
process, leaving out the contribution from heavy boson exchange diagrams.

\par We obtain our numerical results using the standard model of the
electroweak interactions \cite{stanmod}, taking $M_{W}=80.4$ $GeV$ for the
mass of the charged boson $W$ and $M_{Z}=91.2$ $GeV$ for the mass of the
neutral boson $Z$ (hence, $sin^2 \theta_W=0.223$ for the electroweak mixing
angle) \cite{partdata}. We present results for the case of unpolarized deep
inelastic $\nu {\cal N}$-scattering with a neutrino energy in the range
$10^{14} \, eV \leq E_{\nu} \leq 10^{14} \, eV$ and the nucleon at rest$^1$,
$\it{i.e.}$ a target nucleon. We take cuts of 4 $GeV^2$, 4 $GeV^2$ and 10
$GeV^2$ for $Q^2$, $Q'{}^2$ and the invariant mass $W$, respectively. These
values for the cuts are suited for the parton distribution functions of J.
Pumplin {\it et al.} \cite{pumplin} which we will use in our calculations.
In fig. 3, we show the dependence of the cross section for the reaction
$\nu_l + P \rightarrow l^- + W^+ + X$ on the cuts on the momenta transfer
square ($Q^2$, $Q'{}^2$). Taking a cut $W_{c} = 10 \, GeV^2$ for the invariant mass
$W$, we plot $\sigma (\nu_l + P \rightarrow l^- + W^+ + X)$ for the following
cases: a) $Q^2_{c} = Q'{}^2_{c} = 2 \, GeV^2$, b) $Q^2_{c} = Q'{}^2_{c} = 4 \, GeV^2$,
c) $Q^2_{c} = Q'{}^2_{c} = 6 \, GeV^2$ and d) $Q^2_{c} = Q'{}^2_{c} = 8 \, GeV^2$.
We observe in this graph that the total cross section rates of the process
$\nu_l + {\cal N} \rightarrow l^- + W^+ + X$ do not depend strongly on the
choice of the cuts on the momenta transfer square ($Q^2$ and $Q'{}^2$), when
they take on values of a few $GeV^2$.

\par In fig. 4 we plot $\sigma (\nu_l + P \rightarrow l^- + W^+ + X)$ and
$\sigma (\nu_l + P \rightarrow l^- + X)$ as a function of $E_{\nu}$. We
obtain $\sigma ( \nu_l + P \rightarrow l^- + W^+ + X) = 2.3 \times 10^{-35}$
$cm^2$ for $E_{\nu} = 10^{17} \, eV$. In fig. 5, we show $\sigma (\nu_l +
N \rightarrow l^- + W^+ + X)$ and $\sigma (\nu_l + N \rightarrow l^- + X)$
as a function of $E_{\nu}$. We get $\sigma ( \nu_l + N \rightarrow l^- + W^+
+ X) = 1.8 \times 10^{-35}$ $cm^2$ for $E_{\nu} = 10^{17} \, eV$. Finally, in
fig. 6, we present our results for $\sigma (\nu_l + {\cal N} \rightarrow l^-
+ W^+ + X) / \sigma (\nu_l + {\cal N} \rightarrow l^- + X)$ (${\cal N}: P, \,
N$) as a function of $E_{\nu}$, with $E_{\cal N}=m_{\cal N}$.  In particular,
we obtain $\sigma ( \nu_l + P \rightarrow l^- + W^+ + X) / \sigma ( \nu_l + P
\rightarrow l^- + X) = 4.7 \times 10^{-3}$ and $\sigma ( \nu_l + N
\rightarrow l^- + W^+ + X) / \sigma ( \nu_l + N \rightarrow l^- + X) = 3.6
\times 10^{-3}$ for $E_{\nu} = 10^{17} \, eV$.

\section{Conclusions}

\par We have presented the general formulas for the cross section of the
production of massive vector bosons in deep inelastic neutrino nucleon
scattering in the framework of the standard model. The expressions for the
matrix elements are given in a way that is suitable for extended models
containing more vector bosons than the SM.

\par We have neglected the contribution from heavy boson exchange diagrams in
our numerical calculations. Hence, we have given numerical results only for
the total cross section rates of the process $\nu_l + {\cal N} \rightarrow
l^- + W^+ + X$ (${\cal N}$: $P$, $N$.) because this reaction is the only one
which gets contribution from photon-exchange diagrams in the lowest order
of $\alpha$. Considering the nucleon at rest, taking $M_W = 80.4$ $GeV$,
$sin^2 \theta _W=0.223$, a neutrino energy in the range $10^{14} \, eV \leq
E_{\nu} \leq 10^{17} \, eV$, setting cuts of 4 $GeV^2$ and 10 $GeV^2$ for the
momenta transfer square ($Q^2$ and $Q'{}^2$) and the invariant mass ($W$). We
made use of the parton distribution functions of J. Pumplin {\it et al.}, we
used the CTEQ PDFs provided in a $n_f=5$ active flavours scheme. In
particular, we have obtained $\sigma (\nu_l + P \rightarrow l^- + W^+ + X) =
2.3 \times 10^{-35}$ $cm^2$ and $\sigma (\nu_l + N \rightarrow l^- + W^+ + X)
= 1.8 \times 10^{-35}$ $cm^2$ for $E_{\nu} = 10^{17} \, eV$. We have also
shown that the total cross section rates of the process $\nu_l + {\cal N}
\rightarrow l^- + W^+ + X$ do not depend strongly on the choice of the cuts
on the momenta transfer square ($Q^2$ and $Q'{}^2$), when we take them equal
to a few $GeV^2$.

\par We have also presented results for $\sigma (\nu_l + {\cal N} \rightarrow
l^- + W^+ + X) / \sigma (\nu_l + {\cal N} \rightarrow l^- + X)$ (${\cal N}:
P, \, N$) as a function of $E_{\nu}$ in the range $10^{14} \, eV \leq E_{\nu}
\leq 10^{17} \, eV$, with $E_{\cal N}=m_{\cal N}$. We have gotten $\sigma
( \nu_l + P \rightarrow l^- + W^+ + X) / \sigma( \nu_l + P \rightarrow l^-
+ X) = 4.7 \times 10^{-3}$ and $\sigma ( \nu_l + N \rightarrow l^- + W^+ + X)
/ \sigma( \nu_l + N \rightarrow l^- + X) = 3.6 \times 10^{-3}$ for $E_{\nu} =
10^{17} \, eV$.

\begin{center}
{\bf ACKNOWLEDGMENTS}
\end{center}
R. M. G-H. is grateful for the kind hospitality extended to her by the Abdus
Salam International Centre for Theoretical Physics (Trieste, Italy) during
summer stays in 2002 and 2003, when part of this work was made. She also
thanks {\it CONACyT} (M\'exico) and {\it VIEP-BUAP} (Puebla, M\'exico) for
financial support. A. R. would like to acknowledge {\it Sistema Nacional de
Investigadores} and {\it CONACyT} (M\'{e}xico) for financial support.

\newpage

\appendix{}
\section{}

By setting all fermion masses to zero, in this appendix we give the
expressions for the quantities $T^{r,r'}$, which are defined as follows:
\begin{eqnarray*}
T^{r,r'}_{P_q,P_l}={\cal H}_{\mu\mu'}^{P_q}{\cal F}_r^{\mu\nu}
({\cal F}_{r'}^{\mu'\nu'})^* {\cal L}_{\nu\nu'}^{P_l},\hspace{3 cm}
(r,r'=1,2,3),\hspace{2 cm}
\end{eqnarray*} 
with
\begin{eqnarray*}
{\cal L}^{L,R}_{\mu\mu'}&=&2\{p_{\mu} p'_{\mu'}+ p'_{\mu} p_{\mu'}- 
g_{\mu \mu'} pp' 
\mp i\varepsilon_{\mu\mu'\rho\sigma}p^{\rho} p'{}^{\sigma} \} \\
{\cal H}^{L,R}_{\mu\mu'}&=&2\{q_{\mu} q'_{\mu'}+ q'_{\mu} q_{\mu'}-
g_{\mu \mu'} qq'
\mp i\varepsilon_{\mu\mu'\rho\sigma}q^{\rho} q'{}^{\sigma} \} \\
\hspace{7mm} {\cal F}^{\mu \nu}_1&=&g^{\mu \nu} \\
\hspace{7mm} {\cal F}^{\mu \nu}_2&=&\varepsilon^{\mu} k^{\nu} -
\varepsilon^{\nu} k^{\mu} \\
\hspace{7mm} {\cal F}^{\mu \nu}_3&=&i\varepsilon^{\mu \nu \rho \sigma}
\varepsilon_{\rho} k_{\sigma}
\end{eqnarray*} 
and $P_q$, $P_l$ being the polarization of the initial quark and initial
neutrino, respectively. In general, these quantities are functions of scalar
products of the momenta $p$, $q$, $p'$, $q'$ , $k$ and the polarization
vector $\varepsilon$ of the produced boson.

Using the polarization sum for the massive vector boson
\begin{equation}
\sum_{\lambda} \varepsilon_{\mu} (k,\lambda) \varepsilon_{\nu} (k,\lambda)=
-g_{\mu\nu}+\displaystyle \frac{k_{\mu} k_{\nu}}{M^2_B}.
\end{equation} 
and the definitions given in (18) we get
\begin{eqnarray*}
\displaystyle \sum_{\lambda}{\cal F}_2^{\mu\nu}({\cal F}_2^{\rho\sigma})^*&=&-
g^{\mu\rho}\, k^\nu 
k^\sigma+g^{\mu\sigma}\, k^\nu k^\rho+g^{\nu\rho}\, k^\mu k^\sigma-
g^{\nu\sigma}\, k^\mu k^\rho\nonumber\\
\displaystyle \sum_{\lambda}{\cal F}_2^{\mu\nu}({\cal F}_3^{\rho\sigma})^*&=&
i(\varepsilon^{\mu\rho\sigma\beta}k^\nu k_\beta-
\varepsilon^{\nu\rho\sigma\beta}k^\mu k_\beta)\\
\displaystyle \sum_{\lambda}{\cal F}_{3}^{\mu\nu}({\cal F}_{3}^{\rho\sigma})^*
&=&(g^{\mu\rho}
g^{\nu\sigma}-g^{\mu\sigma}g^{\nu\rho})M^2_B+\sum_{\lambda}{\cal F}_2
^{\mu\nu} ({\cal F}_2^{\rho\sigma})^* \nonumber
\end{eqnarray*}
Aided by these expressions we obtain 
\begin{eqnarray*}
\displaystyle \sum_{\lambda}T^{22}&=&\left \{ \begin{array}{llll}
&8[-M^2_B(pp'\cdot qq'+p'q'\cdot pq-p'q\cdot q'p)&\hspace{0.5in} LL,RR\\
&\hspace{3mm} +2(pq\cdot kp'\cdot kq'+p'q'\cdot kp\cdot kq)]&{}\vspace{3mm}\\
&8[-M^2_B(pp'\cdot qq'-p'q'\cdot pq+p'q\cdot q'p)&\hspace{0.5in} LR,RL\\
&\hspace{3mm} +2(pq'\cdot kp'\cdot kq+p'q\cdot kp\cdot kq')]&{}\end{array}
\right. \vspace{3mm}\\
\displaystyle \sum_{\lambda}Re\: T^{23}&=&\left\{ \begin{array}{lll} 
&\pm 16(-pq\cdot kp'\cdot kq'+p'q'\cdot kp\cdot kq)&\hspace{0.7in}  LL,RR
\vspace{3mm} \\
&\mp 16(p'q\cdot kp\cdot kq'-pq'\cdot kp'\cdot kq)&\hspace{0.7in}  LR,RL
\end{array} \right. \vspace{3mm}\\ 
\displaystyle \sum_{\lambda}T^{33}&=&\left \{ \begin{array}{llll}
&8[M^2_B(pp'\cdot qq'-p'q\cdot pq'-pq\cdot p'q')&\hspace{0.5in}  LL,RR\\
&\hspace{3mm} +2(pq\cdot kp'\cdot kq'+p'q'\cdot kp\cdot kq)]&{}\vspace{3mm}\\
&8[M^2_B(pp'\cdot qq'-p'q\cdot pq'-pq\cdot p'q')&\hspace{0.5in} LR,RL\\
&\hspace{3mm} +2(pq'\cdot kp'\cdot kq+p'q\cdot kp\cdot kq')]&{}\end{array}
\right.
\end{eqnarray*}
The remaining $T$,s can be expressed as follows
\begin{eqnarray*}
T^{11}&=&\left\{ \begin{array}{lll}
&16p'q'\cdot pq&\hspace{1.9in}  LL,RR \vspace{3mm} \\
&16p'q\cdot pq'&\hspace{1.9in} LR,RL\end{array} \right. \vspace{3mm}\\
Re\: T^{12}&=&\left\{ \begin{array}{llll} &-8(\varepsilon p\cdot p'q'\cdot qk
+\varepsilon p'\cdot pq\cdot q'k&\hspace{0.5in} LL,RR\\
&\hspace{0.5cm} -\varepsilon q\cdot p'q'\cdot pk-\varepsilon q'\cdot pq\cdot p'k)&{}
\vspace{3mm} \\
&-8(\varepsilon p\cdot p'q\cdot q'k+\varepsilon p'\cdot pq'\cdot qk&\hspace{0.5in}
 LR,RL\\
&\hspace{0.5cm} -\varepsilon q\cdot pq'\cdot p'k-\varepsilon q'\cdot p'q\cdot pk&{}
\end{array} \right. \vspace{3mm}\\
Re\:  T^{13}&=&\left\{ \begin{array}{llll} &\mp 8(\varepsilon p\cdot p'q'
\cdot qk -\varepsilon p'\cdot pq\cdot q'k&\hspace{0.5in} LL,RR \\
&\hspace{0.5cm} +\varepsilon q'\cdot pq\cdot p'k-\varepsilon q\cdot p'q'\cdot pk)&{}
\vspace{3mm} \\
&\pm 8(\varepsilon p\cdot p'q\cdot q'k-\varepsilon p'\cdot pq'\cdot qk&
\hspace{0.5in} LR,RL\\
&\hspace{0.5cm} -\varepsilon q'\cdot p'q\cdot pk+\varepsilon q\cdot pq'\cdot
p'k)&{} \end{array} \right.
\end{eqnarray*}
Using (A1) and (18) we get
\begin{eqnarray*}
\sum_{\lambda}\varepsilon P_1\cdot \varepsilon P_2&=&-P_1P_2+\displaystyle 
\frac{kP_1\cdot kP_2}{M^2_B}\\ 
\sum_{\lambda}\varepsilon P\cdot {\cal F}^{\mu \nu}_2&=&{\cal F}^{\mu 
\nu}_2(\varepsilon\rightarrow -P)\\
\sum_{\lambda}\varepsilon P\cdot {\cal F}^{\mu \nu}_3&=&{\cal F}^{\mu \nu}_
3(\varepsilon\rightarrow -P)
\end{eqnarray*}
for $P, \, P_1, \, P_2=p,\, q, \, p', \, q'$. Hence
\begin{eqnarray*}
\sum_{\lambda}\varepsilon P_1\cdot \varepsilon P_2 \, T^{11}&=&(-P_1P_2+
\displaystyle \frac{kP_1\cdot kP_2}{M^2_B})\, T^{11}\\ 
\sum_{\lambda}\varepsilon P\cdot Re\: T^{12}&=&Re\: T^{12}(\varepsilon
\rightarrow -P)\\
\sum_{\lambda}\varepsilon P\cdot Re\:  T^{13}&=&Re\: T^{13}(\varepsilon
\rightarrow -P)
\end{eqnarray*}

\newpage

\begin{center}
{\bf Figure Captions}
\end{center}

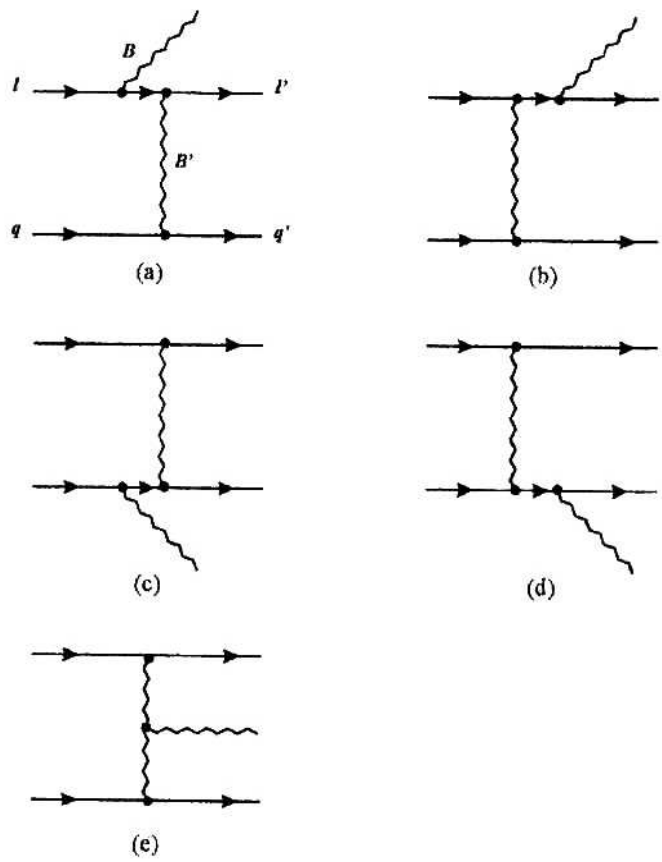
\noindent{\bf Fig. 1}: Feynman diagrams for heavy boson production
from the initial (a) and final (b) lepton, from the initial (c) and final (d)
quark and via the non-Abelian couplings (e).

\noindent{\bf Figs. 2.1 - 2.5}: Feynman diagrams for processes (P.1) - (P.5):
boson production from the incoming neutrino (a), the outgoing lepton (b), the
initial (c) and final (d) quark, and through the non-Abelian couplings (e).
$u$ stands for $u$, $c$, $\bar d$, $\bar s$, $\bar b$; $d$ for $d$, $s$,
$b$, $\bar u$, $\bar c$.

\bigskip
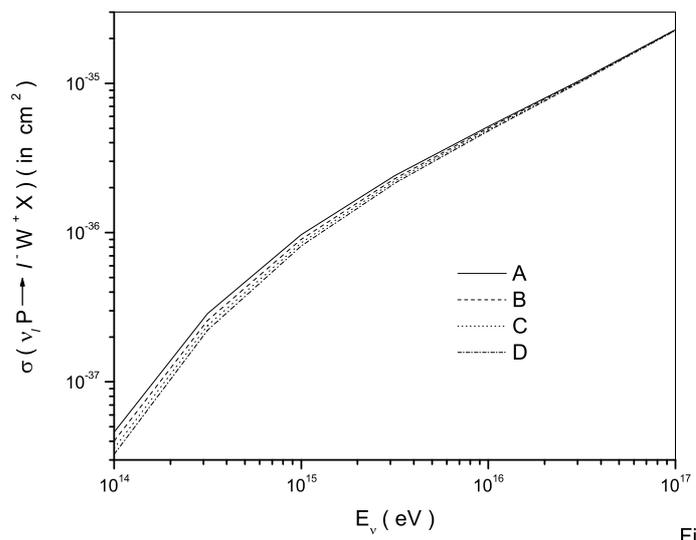
\noindent{\bf Fig. 3}:
$\sigma (\nu_l + P \rightarrow l^- + W^+ + X)$ as a function of $E_{\nu}$ in
the range $10^{14} \, eV \leq E_{\nu} \leq 10^{17} \, eV$, with $E_P=m_P$.
Taking $W_{c} = 10 \, GeV^2$ and a) $Q^2_{c} = Q'{}^2_{c} = 2 \, GeV^2$
(curve A), b) $Q^2_{c} = Q'{}^2_{c} = 4 \, GeV^2$ (curve B), c) $Q^2_{c} =
Q'{}^2_{c} = 6 \, GeV^2$ (curve C), d) $Q^2_{c} = Q'{}^2_{c} = 8 \, GeV^2$
(curve D).

\bigskip

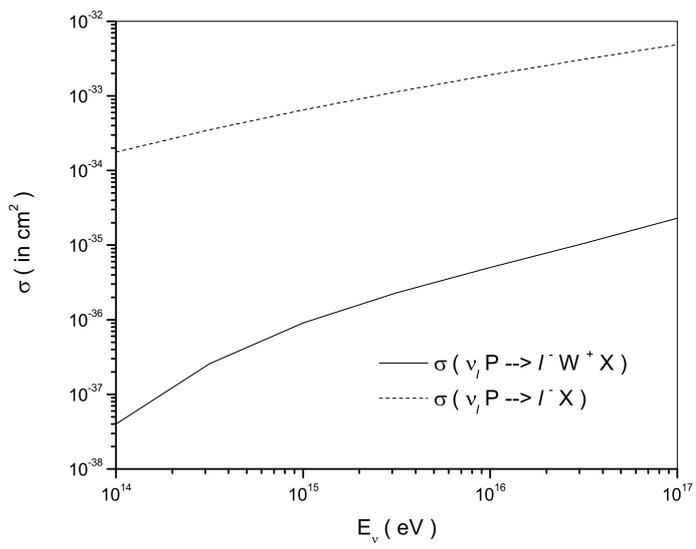
\noindent{\bf Fig. 4}:
$\sigma (\nu_l + P \rightarrow l^{-} + W^{+} + X)$ and $\sigma (\nu_l + P
\rightarrow l^{-} + X)$ as a function of $E_{\nu}$ in the range
$10^{14} \, eV \leq E_{\nu} \leq 10^{17} \, eV$, with $E_P=m_P$.
Taking $W_{c} = 10 \, GeV^2$ and $Q^2_{c} = Q'{}^2_{c} = 4 \, GeV^2$.

\bigskip

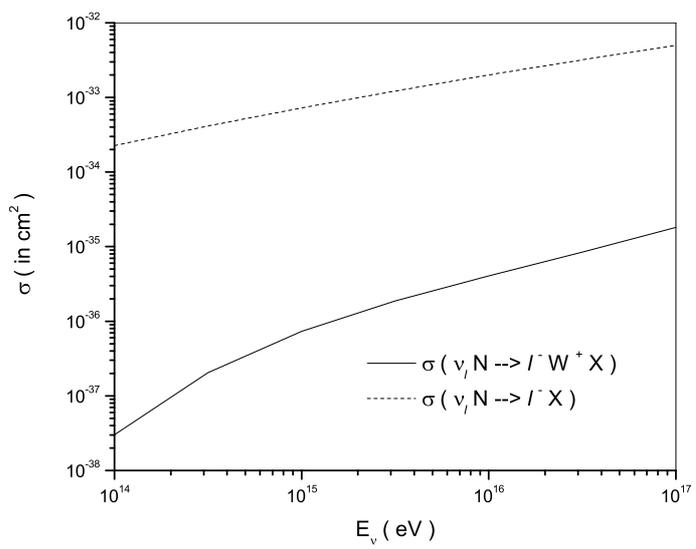
\noindent{\bf Fig. 5}:
$\sigma (\nu_l + N \rightarrow l^{-} + W^{+} + X)$ and $\sigma (\nu_l + N
\rightarrow l^{-} + X)$ as a function of $E_{\nu}$ in the range
$10^{14} \, eV \leq E_{\nu} \leq 10^{17} \, eV$, with $E_N=m_N$.
Taking $W_{c} = 10 \, GeV^2$ and $Q^2_{c} = Q'{}^2_{c} = 4 \, GeV^2$.

\bigskip

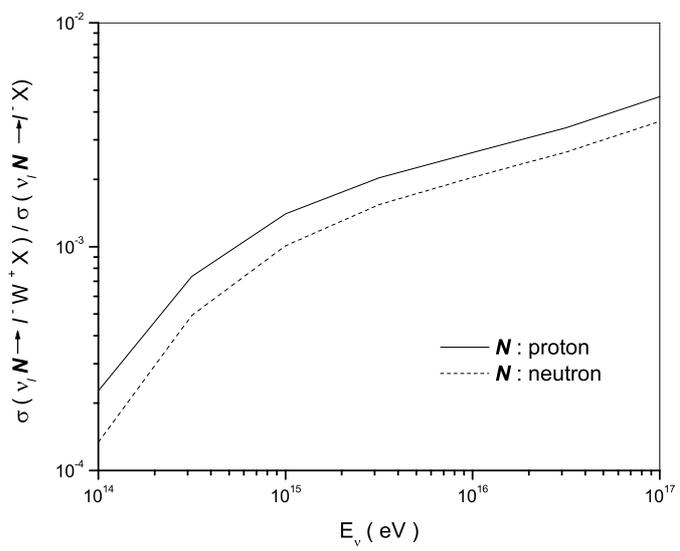
\noindent{\bf Fig. 6}:
$\sigma (\nu_l + {\cal N} \rightarrow l^{-} + W^{+} + X) / \sigma (\nu_l +
{\cal N} \rightarrow l^{-} + X)$ (${\cal N}: P, \, N$) as a function of
$E_{\nu}$ in the range $10^{14} \, eV \leq E_{\nu} \leq 10^{17} \, eV$,
with $E_{\cal N}=m_{\cal N}$.
Taking $W_{c} = 10 \, GeV^2$ and $Q^2_{c} = Q'{}^2_{c} = 4 \, GeV^2$.

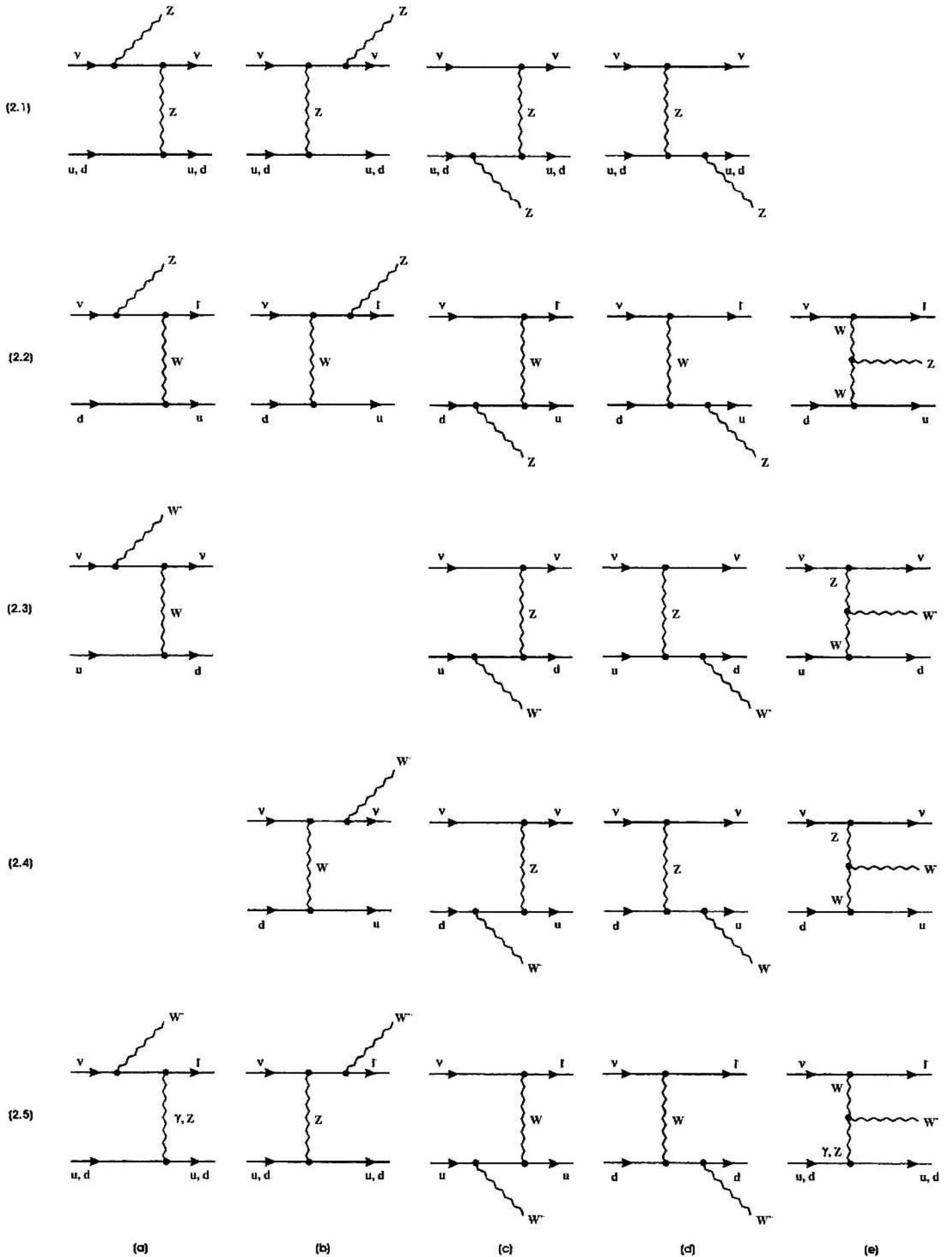
\end{document}